\documentclass[12pt, a4paper]{article}
\usepackage{epsf}
\usepackage{cite}
\usepackage{amsmath,amssymb}
\input{colordvi.tex}
\usepackage[usenames,dvipsnames]{color}
\usepackage{graphicx}
\usepackage{ascmac}
\usepackage{hyperref}
\usepackage{slashed}

\begin{document}

\begin{titlepage}

\begin{center}
\hfill TU-1253\\
\hfill KEK-QUP-2024-0029
\vskip 1.in

\renewcommand{\thefootnote}{\fnsymbol{footnote}}

{\Large \bf
Domain walls in Nelson-Barr axion model
}

\vskip .5in

{\large
Kai Murai$^{(a)}$\footnote{kai.murai.e2@tohoku.ac.jp}
and
Kazunori Nakayama$^{(a,b)}$\footnote{kazunori.nakayama.d3@tohoku.ac.jp}
}

\vskip 0.5in

$^{(a)}${\em 
Department of Physics, Tohoku University, 6-3 Aoba, Aramaki, Aoba-ku, Sendai, Miyagi 980-8578, Japan
}

\vskip 0.2in

$^{(b)}${\em 
International Center for Quantum-field Measurement Systems for Studies of the Universe and Particles (QUP), KEK, 1-1 Oho, Tsukuba, Ibaraki 305-0801, Japan
}

\end{center}
\vskip .5in

\begin{abstract}
We explore a concrete realization of a Nelson-Barr model addressing the strong CP problem with suppressed unfavorable corrections.
This model has a scalar field that spontaneously breaks discrete symmetry, and its phase component can naturally be relatively light, which we call the Nelson-Barr axion.
It has both a tree-level potential and the QCD instanton-induced potential like the QCD axion, each minimizing at the CP-conserving point.
While one potential leads to domain wall formation, the other works as a potential bias.
This model provides a natural setup for the collapse of the axion domain walls by a potential bias without spoiling a solution to the strong CP problem.
We discuss the cosmological implications of domain wall collapses, including dark matter production and gravitational wave emission.
\end{abstract}

\end{titlepage}

\tableofcontents

\renewcommand{\thefootnote}{\arabic{footnote}}
\setcounter{footnote}{0}

\section{Introduction}

The strong CP problem remains one of the unresolved issues in the Standard Model (see, e.g., Refs.~\cite{Kim:2008hd,Hook:2018dlk} for reviews).
Quantum chromodynamics (QCD) generally includes a CP-violating term $\propto \theta_s \mathrm{Tr}[G_{\mu\nu} \tilde{G}^{\mu\nu}]$ with $G_{\mu\nu}$ and $\tilde{G}^{\mu\nu}$ being the field strength of the gluon and its dual, respectively.
Considering the chiral rotation of the quark phases, the CP violation is described by an invariant angle of
\begin{align}
    \bar{\theta}_s 
    \equiv
    \theta_s + \arg [\det ({\textbf m}_u {\textbf m}_d)]
    \ ,
\end{align}
where ${\textbf m}_u$ and ${\textbf m}_d$ are the mass matrices of the up- and down-type quarks, respectively.
From the perspective of naturalness, $\bar{\theta}_s$ is expected to be of $\mathcal{O}(1)$.
On the other hand, the measurement of the neutron electric dipole moment imposes a stringent limit of $|\bar{\theta}_s| \lesssim 10^{-10}$~\cite{Abel:2020pzs}.
This discrepancy is called the strong CP problem, and various solutions have been suggested.
Among them, there are two well-known classes of solutions to the strong CP problem:%
\footnote{
    The solution with the massless up-quark~\cite{Georgi:1981be,Kaplan:1986ru,Choi:1988sy,Banks:1994yg} is currently disfavored by the results of lattice QCD~\cite{Fodor:2016bgu,Alexandrou:2020bkd,FlavourLatticeAveragingGroupFLAG:2021npn}.
}
the Peccei-Quinn mechanism~\cite{Peccei:1977hh,Peccei:1977ur} with the QCD axion~\cite{Weinberg:1977ma,Wilczek:1977pj} and the spontaneous or soft breaking of a discrete spacetime symmetry such as parity~\cite{Beg:1978mt,Mohapatra:1978fy,Babu:1989rb,Barr:1991qx} or CP symmetry~\cite{Georgi:1978xz,Segre:1979dx,Barr:1979as,Nelson:1983zb,Barr:1984qx,Bento:1991ez}.

In the Nelson-Barr mechanism~\cite{Nelson:1983zb,Barr:1984qx}, one realization of the latter idea, the fundamental theory is exactly symmetric under a CP transformation, and thus $\bar{\theta}_s = 0$ at the Lagrangian level.
With a specific structure of the mass matrix, the CP-violating angle in the Cabibbo-Kobayashi-Maskawa (CKM) matrix can be reproduced while keeping the smallness of $\bar{\theta}_s$.
The class of the Nelson-Barr model and its extensions remain an active field of research even in this decades (see, e.g., Refs.~\cite{Vecchi:2014hpa,Dine:2015jga,Davidi:2017gir,Schwichtenberg:2018aqc,Cherchiglia:2019gll,Evans:2020vil,Cherchiglia:2020kut,Perez:2020dbw,Cherchiglia:2021vhe,Valenti:2021rdu,Valenti:2021xjp,Fujikura:2022sot,Girmohanta:2022giy,McNamara:2022lrw,Asadi:2022vys,Perez:2023zin,Suematsu:2023jqa,Banno:2023yrd,Dine:2024bxv,Bastos:2024afz}).

Recently, we proposed a modification of the minimal realization of the Nelson-Barr mechanism.
Compared to the original model proposed by Bento, Branco, and Parada (BBP)~\cite{Bento:1991ez}, we introduced an approximate discrete symmetry, $Z_{4n}$, to the model. 
Thanks to this symmetry, unfavorable contributions to $\bar{\theta}_s$ from loop diagrams and higher-order operators are highly suppressed by a small parameter $\epsilon$.
As a result, we can safely realize a high scale of spontaneous CP breaking $\gg 10^8$\,GeV, which is compatible with the required temperature $T \gtrsim 10^9$\,GeV~\cite{Giudice:2003jh,Buchmuller:2004nz} for the thermal leptogenesis scenario~\cite{Fukugita:1986hr}.
While the previous work investigated the requirement for the model parameters including $\epsilon$, the model was presented as an effective theory, and its concrete realization was not discussed in detail.

In this paper, we propose one possible realization of the Nelson-Barr mechanism in Ref.~\cite{Murai:2024alz}.
The soft breaking of the additional symmetry is described by the vacuum expectation value (VEV) of a complex scalar $\phi$ with a charge of $Z_{4n}$.
While the VEV of $\phi$ must be real to solve the strong CP problem, $\phi$ necessarily has multiple potential minima including complex ones due to the $Z_{4n}$ symmetry.
Thus, cosmic strings and domain walls can be formed if $\phi$ acquires nonzero VEV after inflation, and $\bar{\theta}_s$ is unfavorably large in all domains but $\mathrm{Im}\phi = 0$.
In this sense, the strong CP problem is solved with a probability of $1/(4n)$.
Interestingly, the introduction of the complex scalar necessarily predicts something like the QCD axion.
Since the complex scalar couples to quarks, the QCD non-perturbative effects induce a potential for its phase component. If the QCD potential is dominant over the tree-level potential, it is nothing but a QCD axion. 
In our case, the strong CP problem is solved via the Nelson-Barr mechanism even if the axion potential is dominated by the tree level one. We call it the Nelson-Barr axion.

We study the cosmology and phenomenology of the Nelson-Barr axion model. What we find are:
\begin{itemize}
\item Compared with the conventional QCD axion models, the Nelson-Barr axion allows different parameter regions in terms of the mass and coupling strength.
\item Domain walls are formed when the $Z_{4n}$ symmetry is spontaneously broken. Such walls are unstable due to the potential bias given by the QCD effect. It may leave the observable gravitational wave signals.
\item Nelson-Barr axion can play the role of dark matter, either from the misalignment mechanism or the collapse of axionic domain walls.
\end{itemize}
The remarkable point of this model is that the collapse of axion domain walls can be realized without spoiling the quality of the QCD axion.%
\footnote{
    The effect of the QCD potential as a bias on the domain wall has been discussed in Refs.~\cite{Preskill:1991kd,Dine:1993yw,Riva:2010jm,Moroi:2011be,Hamaguchi:2011nm,Higaki:2016yqk,Higaki:2016jjh,Chigusa:2018hhl,Chiang:2020aui,Ferreira:2022zzo,Kitajima:2023cek,Bai:2023cqj,Blasi:2023sej}.
}
This is in contrast to simple axion models, where an additional source of the axion potential for the domain wall collapse generally spoils the quality of the QCD axion unless a fine-tuning of the potential is assumed.

The rest of this paper is organized as follows. 
In Sec.~\ref{sec: model}, we briefly review the modified BBP model proposed in Ref.~\cite{Murai:2024alz} and provide its concrete realization. We also discuss the emergence of the Nelson-Barr axion.
In Sec.~\ref{sec: domain wall}, we investigate the formation, evolution, and collapse of cosmic strings and domain walls and evaluate the axions and gravitational waves emitted from domain walls.
Finally, we conclude and discuss our results in Sec.~\ref{sec: conclusion}.

\section{Realization of Nelson-Barr model and axion}
 \label{sec: model}

\subsection{Modified BBP model}

First, we review the model in the Nelson-Barr framework proposed in Ref.~\cite{Murai:2024alz}, which is an extension of a minimal model of spontaneous CP violation~\cite{Bento:1991ez}.
In this model, we impose CP symmetry and introduce vector-like quarks $D_L$ and $D_R$ with the U(1) hypercharge of $-1/3$, right-handed neutrinos $N_i$, and a singlet complex scalar $S$ as additional field contents to the Standard Model.
The Lagrangian is given by 
\begin{align}
    \mathcal L 
    =&
    -\Big[ 
        \epsilon^k M\overline D_L D_R 
        + \epsilon^k (g_i S + g_i' S^*) \overline D_L d_{Ri} 
        + y^d_{ij} H \overline Q_{Li} d_{Rj} 
    \nonumber \\
    & ~~~~ 
        + \frac{1}{2} (\xi_{ij} S + \xi_{ij}' S^*) \overline{N^c_i} N_j  
        + y_{ij}^\nu \widetilde H \overline L_i N_j 
        + y_{ij}^e H \overline L_i e_{Rj}
        + {\rm h.c.}
    \Big] 
    - V(S,H)
    \ ,
\end{align}
where $\epsilon$, $M$, $g_i$, $g_i'$, $y_{ij}^d$, $\xi_{ij}$, $\xi_{ij}'$, $y_{ij}^\nu$, $y_{ij}^e$ $(i,j=1,2,3)$ are real constants due to the CP symmetry, and $d_{Ri}, Q_{L_i}, H, L_i, e_{Ri}$ are the Standard Model right-handed down-type quarks, left-handed quark doublets, Higgs doublet, left-handed lepton doublets, and right-handed leptons, respectively.
Note that the right-handed neutrinos are introduced to accommodate leptogenesis and are not essential to solve the strong CP problem.
Here, we impose an exact $Z_4$ symmetry and an approximate $Z_{4n}^\mathrm{(app)}$ symmetry and assign their charges as shown in Table~\ref{tab: charge assignment}.
While the former symmetry forbids the interaction terms such as $H \overline Q_{Li} D_R$, the latter forbids the terms with $\epsilon$, which can be regarded as a spurion field with a $\pm 1$ charge under the $Z_{4n}^{\rm (app)}$ symmetry.
We consider that $Z_{4n}^\mathrm{(app)}$ is softly broken, and $\epsilon^k \ll 1$.
Due to the smallness of $\epsilon^k$, this model is free from the quality problem in the Nelson-Barr models.
\begin{table}[t]
    \centering
    \begin{tabular}{|c|c|c|c|c|c|c|c|c|c|c|} 
    \hline
    ~ & $S$ & $D_L$ & $D_R$ & $Q_{Li}$ & $d_{Ri}$ & $u_{Ri}$ & $H$ & $N_i$ & $L_i$ & $e_{Ri}$ 
    \\ \hline
    $Z_4$  & $2$ & $2$ & $2$ & $0$ & $0$ & $0$ & $0$ & $1$ & $1$ & $1$ 
    \\ \hline
    $Z^{\rm (app)}_{4n}$  & $2n$ & $2n$ & $2n-k$ & $k$ & $k$ & $k$ & $0$ & $n$ & $n$ & $n$
    \\ \hline
    \end{tabular}
    \caption{Charge assignments on the fields in the modified BBP model. }
    \label{tab: charge assignment}
\end{table}

The scalar potential is given by
\begin{align}
    V(S,H) 
    =&
    \lambda_H\left(|H|^2-v_H^2\right)^2 
    +  \lambda_S\left(|S|^2-v_S^2\right)^2 
    - \lambda_{SH}\left(|H|^2-v_H^2\right)\left(|S|^2-v_S^2\right) 
    \nonumber \\ 
    &+ (\mu^2 + \gamma_{SH} |H|^2 + \gamma_{2}|S|^2)(S^2+S^{*2}) 
    + \gamma_4 (S^4+ S^{*4})
    \ ,
\end{align}
where all coefficients are real due to the CP symmetry, and $v_H = 174$\,GeV
and $v_S$ are the VEVs of $H$ and $S$, respectively.
For simplicity, we assume that the terms in the second line are smaller than those in the first line so that the VEVs of $|H|$ and $|S|$ are mostly determined by the first line.
With this potential, the VEV of $S$ is given by $S = v_S e^{i\theta_S}$, where the complex phase $\theta_S$ can take an arbitrary value depending on the choice of coefficients of the $S^2+S^{*2}$ and $S^4+S^{*4}$ terms. This is the source of spontaneous CP violation.

If $S$ acquires the nonzero VEV after inflation, domain walls are formed at the spontaneous symmetry breaking and may alter the expansion history of the later universe.
We can avoid such a situation by requiring a low reheating temperature so that the symmetry is not restored after inflation. 
Instead, we can assure that the symmetry remains spontaneously broken after inflation by assuming negative Hubble and thermal mass potentials for $S$ with $\gamma_{SH} < 0$~\cite{Murai:2024alz}.
Note that the latter approach allows a high reheating temperature, which is required for thermal leptogenesis.

Then, the mass matrix of the quarks can be written as
\begin{align}
    -\mathcal L 
    =
    (\overline d_{Li}, \overline D_L) \mathcal M 
    \begin{pmatrix}
        d_{R j} \\ D_R	
    \end{pmatrix} 
    + {\rm h.c.},
    ~~~~~~
    \mathcal M 
    =
    \begin{pmatrix}
        {\textbf m}_{ij} &  0 \\ B_j  & \epsilon^k M	
    \end{pmatrix},
    \label{eq: mass matrix}
\end{align}
where ${\textbf m}_{ij} \equiv y^d_{ij} v_H$ and $B_j \equiv \epsilon^k (g_j e^{i\theta} + g_j' e^{-i\theta} ) v_S$. 
Here, $B_i$ is complex while ${\textbf m}_{ij}$ and $\epsilon^k M$ are real.
Thus, $\arg [ \det \mathcal{M} ] = 0$, and the strong CP angle vanishes at the tree level.
On the other hand, the CP phase in the CKM matrix appears after diagonalizing this mass matrix.
To reproduce $\mathcal{O}(1)$ CP phase in the CKM matrix while keeping the unitarity of the CKM matrix, we require $|\epsilon^k M| \sim |B_j|$ and $|B{\textbf m}^\mathrm{T}|^2 / |\epsilon^k M|^4 \ll 1$~\cite{Murai:2024alz}.

When we take into account higher-order operators and loop effects, there are additional contributions to the strong CP angle.
One possible operator contributing to the strong CP angle is given by 
\begin{align}
    \mathcal L 
    =
    \epsilon^{2k} \frac{c_iS + c_i'S^*}{\Lambda} H \overline Q_{Li} D_R 
    + {\rm h.c.},
    \label{eq: higher dimensional operators}
\end{align}
with cutoff scale $\Lambda$ and $c_i, c_i'$ being real constants. 
Due to the $Z_{4n}^\mathrm{(app)}$ symmetry, this operator is suppressed by $\epsilon^{2k}$.
Furthermore, there are loop contributions to the strong CP angle, which is estimated as~\cite{Murai:2024alz}
\begin{align}
    \bar\theta_s 
    \sim 
    \frac{\epsilon^{2k}}{32\pi^2} \gamma_{SH}\sin(2\theta)\sum_i(g_i^2 -g_i'^2) 
    \log\left( \frac{m_h^2}{m_\sigma^2} \right)
    \ ,
\end{align}
where $m_h$ and $m_\sigma$ are the masses of the Higgs particle and the angular component of $S$, respectively.
These additional contributions to $\bar{\theta}_s$ are suppressed by $\epsilon^{2k}$ and are safely small, $\lesssim 10^{-10}$, with sufficiently small $\epsilon^{2k}$ even if the model parameters such as $v_S/\Lambda$, $\gamma_{SH}$, $g_i$, and $g_i'$ are not tuned.
This is an advantage of the modified model~\cite{Murai:2024alz} compared to the original setup~\cite{Bento:1991ez}.

\subsection{Concrete realization of the modified BBP model}

\subsubsection{Model description} \label{sec:modeldesc}

So far, the small parameter $\epsilon$ is an effective parameter that describes the small explicit breaking of the $Z_{4n}$ symmetry.
Next, we provide a concrete realization of this model. 
We assume that the $Z_{4n}$ symmetry is not explicitly broken but only spontaneously broken by a complex scalar $\phi$.
The Lagrangian is given by 
\begin{align}
    \mathcal L 
    =&
    -\left[ 
        \left( \frac{\phi}{\Lambda} \right)^k M 
        \overline D_L D_R 
        + \left( \frac{\phi^*}{\Lambda} \right)^k (g_i S + g_i' S^*) \overline D_L d_{Ri} 
        + y^d_{ij} H \overline Q_{Li} d_{Rj} 
        + {\rm h.c.}
    \right] 
    - V(S,H)
    - V_\phi(\phi)
    \ ,
\end{align}
where we omitted the lepton sector, which is the same as in the previous subsection.
Again all coefficients are real since the CP symmetry is imposed.
We assume that the mass parameter $M$ is of the same order as the mass parameters in the scalar potential and is much smaller than the cutoff scale $\Lambda$.
The charges of the fields are summarized in Table~\ref{tab: charge assignment 2}.
The potential for $\phi$ is given by
\begin{align}
    V_\phi(\phi)
    =
    \lambda_\phi \left(|\phi|^2 - \frac{f_a^2}{2} \right)^2
    - \frac{\gamma_\phi}{\Lambda^{4n-4}} (\phi^{4n} + \phi^{* 4n})
    \ ,
\end{align}
where $\lambda_\phi$ and $\gamma_\phi$ are real constants, and $v_\phi \equiv f_a/\sqrt{2}$ is the VEV of $|\phi|$. 
When $\phi$ has a real VEV, we can regard $\epsilon = v_\phi/\Lambda$ in the previous model.\footnote{
    A similar idea in the context of flavor symmetry to explain the hierarchical Yukawa structure is known as the Froggatt-Nielsen mechanism~\cite{Froggatt:1978nt}.
}
The second term explicitly breaks a U(1) symmetry associated with $\phi$, and ensures that $\phi$ has a potential minimum on its real axis.
We assume $f_a \ll \Lambda$, and then the potential minimum is given by 
\begin{align}
    \phi = \frac{f_a}{\sqrt{2}} e^{i \theta_\phi}
    \ ,
\end{align}
with
\begin{align}
    \theta_\phi 
    =
    \frac{l \pi}{2n} ,
    \quad
    (l = 0, \ldots, 4n-1) .
\end{align}
The strong CP angle is zero at the tree level if we live in the universe with $l=0$. Otherwise, there appears a large strong CP angle inconsistent with experiments. 
In this sense, the solution to the strong CP problem may work with a probability of $1/(4n)$, if we start with one of these vacua during inflation. However, as we will see below, there is a dynamical mechanism to choose the CP-conserving vacuum $l=0$.

Let us comment on the ``quality'' of the CP conservation at the vacuum $l=0$. 
Actually, the reality of the VEV of $\phi$ should not be exact, but there are several terms that cause a deviation from $\theta_\phi=0$. One possible term is $V \sim \phi^{8n} + \phi^{*8n}$, but it does not cause such a deviation if the coefficient of this term is negative.
Moreover, even if the coefficient is positive, it is suppressed by $\epsilon^{4n}$ compared with the original $\phi^{4n} + \phi^{*4n}$ term, and hence the effect of this term is negligible. 
On the other hand, terms such as $V \sim S^2 \phi^{4n} + {\rm h.c.}$ cause deviation from $\theta_\phi=0$; we may have $\theta_\phi \sim \epsilon^2 \sin(2\theta_S)/(4n)$ at the true minimum. This is of the same order as the loop-induced strong CP angle, and it gives a strong constraint on our model parameters.
\begin{table}[t]
    \centering
    \begin{tabular}{|c|c|c|c|c|c|c|c|c|c|c|c|} 
    \hline
    ~ & $\phi$ & $S$ & $D_L$ & $D_R$ & $Q_{Li}$ & $d_{Ri}$ & $u_{Ri}$ & $H$ & $N_i$ & $L_i$ & $e_{Ri}$
    \\ \hline
    $Z_4$  & $0$ & $2$ & $2$ & $2$ & $0$ & $0$ & $0$ & $0$ & $1$ & $1$ & $1$
    \\ \hline
    $Z_{4n}$ & $1$ & $2n$ & $2n$ & $2n-k$ & $k$ & $k$ & $k$ & $0$ & $n$ & $n$ & $n$
    \\ \hline
    \end{tabular}
    \caption{Charge assignments on the fields in the concrete realization of the modified BBP model. }
    \label{tab: charge assignment 2}
\end{table}

\begin{figure}[t]
    \centering
    \includegraphics[width=.8\textwidth]{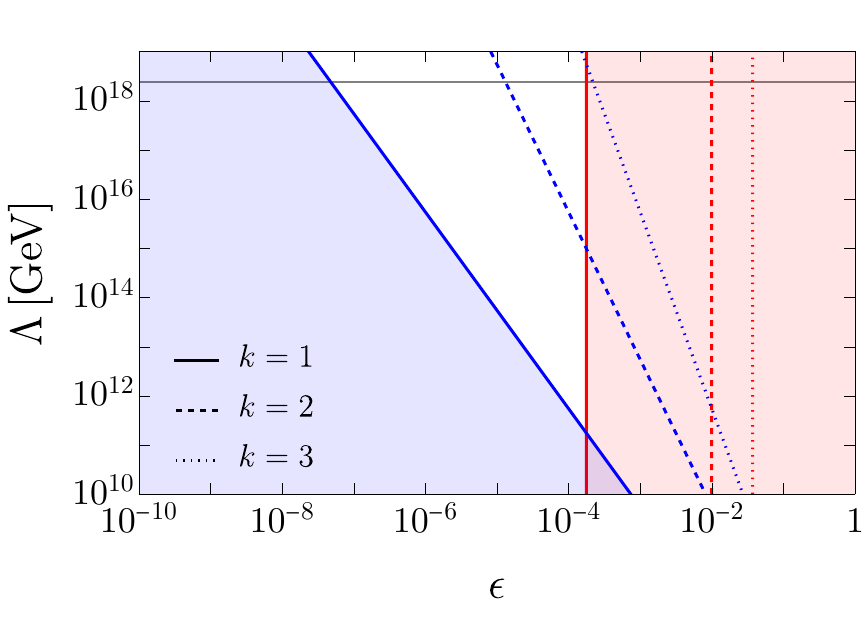}
    \caption{%
        Favored range of $\epsilon$ depending on $\Lambda$ and $k$.
        The blue and red lines denote the lower and upper limits on $\epsilon$, respectively.
        The solid, dashed, and dotted lines correspond to $k = 1$, $2$, and $3$, respectively.
        In the colored regions, $\epsilon$ does not satisfy the condition for $k = 1$.
        The horizontal gray line denotes $\Lambda = M_\mathrm{Pl}$.
    }
    \label{fig: epsilon range}
\end{figure}

Here we summarize the conditions for the model parameter following Ref.~\cite{Murai:2024alz}.
For simplicity, we assume $M = v_S = v_\phi$ and $(g_j e^{i\theta} + g_j' e^{-i\theta} ) = \mathcal{O}(1)$ in the following.
To solve the strong CP problem, $\epsilon = f_a/(\sqrt{2}\Lambda)$ should be sufficiently small: $\epsilon^{2k}/(32\pi^2) \lesssim 10^{-10}$ and $\epsilon^{2k} v_S/\Lambda \lesssim 10^{-10}$.
In addition, to reproduce the CKM matrix, we require $f_a \sim (g_j e^{i\theta} + g_j' e^{-i\theta} ) v_S$ and $v_H^2/v_S^2 \lesssim 10^{-3} \epsilon^{2k}$.
Then, we obtain the conditions of 
\begin{align}
    (4.8
    \times 10^{-8})^{\frac{2}{k+1}} \left( \frac{\Lambda}{M_\mathrm{Pl}} \right)^{-\frac{1}{k+1}}
    \lesssim
    \epsilon 
    \lesssim
    \left\{
        \begin{array}{ll}
            1.8 \times 10^{-4} & \quad (k = 1)
            \\
            (4.6 \times 10^{-4})^{\frac{3}{2k+1}} & \quad (k \geq 2)
        \end{array}
    \right.
    \ ,
    \label{eq: epsilon range}
\end{align}
where $M_\mathrm{Pl} \simeq 2.4 \times 10^{18}$\,GeV is the reduced Planck mass.
We show the favored range of $\epsilon$ in Fig.~\ref{fig: epsilon range}.
For example, when $\Lambda = M_\mathrm{Pl}$ and $ k = 1$, the constraint on $\epsilon$ is given by 
\begin{align}
    4.8
    \times 10^{-8}
    \lesssim
    \epsilon 
    \lesssim
    1.8 \times 10^{-4}
    \ .
    \label{eq: epsilon range simple}
\end{align}

\subsubsection{Appearance of Nelson-Barr axion}

In our model we have a complex scalar $\phi$ that couples to quarks. Thus its angular component obtains mass from the QCD effect.
As a canonical degree of freedom, we define $a \equiv f_a \theta_\phi$. 
Then, non-perturbative effects of the QCD induce a potential of
\begin{align}
    V_\mathrm{QCD}
    =
    \chi(T) \left[  1 - \cos \left( k \frac{a}{f_a} \right) \right]
    \ ,
\end{align}
where $\chi(T)$ is the topological susceptibility of the QCD and depends on temperature as
\begin{align}
    \chi(T)
    \simeq
    \left\{
        \begin{array}{ll}
            \chi_0 & \quad (T < T_\mathrm{QCD})
            \\
            \chi_0 \left( \frac{T}{T_\mathrm{QCD}} \right)^{-p} & \quad (T \geq T_\mathrm{QCD})
        \end{array}
    \right.
    \ .
\end{align} 
In the following, we adopt $\chi_0 = (75.6\,\mathrm{MeV})^4$, $T_\mathrm{QCD} = 153\,\mathrm{MeV}$, and $p = 8.16$~\cite{Borsanyi:2016ksw}.
Taking into account the tree-level term in $V_\phi$, the total potential for $a$ is given by 
\begin{align}
    V_a
    =
    \chi(T) \left[  1 - \cos \left( k \frac{a}{f_a} \right) \right]
    +
    \frac{m_a^2 f_a^2}{16n^2} \left[  1 - \cos \left( 4n \frac{a}{f_a} \right) \right]
    \ ,
    \label{eq: axion potential}
\end{align}
where $m_a^2 \equiv 2^{5-2n} n^2 \gamma_\phi f_a^{4n-2}/\Lambda^{4n-4} = 16 n^2 \gamma_\phi \epsilon^{4n-2} \Lambda^2 $.
If the QCD-induced potential is dominant over the tree-level potential, $a$ may be identified with the QCD axion. In such a case, we do not need to impose the CP symmetry to solve the strong CP problem. However, in our model, the strong CP problem is solved via the Nelson-Barr mechanism even if the QCD-induced potential is subdominant. In this sense, we do not require the ``quality'' of the approximate U(1) symmetry at the level of conventional QCD axion models. We call $a$ the Nelson-Barr axion in such a case. It leads to rich phenomenology.%
\footnote{
    Ref.~\cite{Dine:2024bxv} also considered axion-like particle in the Nelson-Barr model. In their model the axion is a phase of $S$ field by imposing U(1) symmetry with high quality and it is anomaly-free under the QCD.
}

We briefly comment on the cosmology of the present model.
First let us consider a scenario where the $Z_{4n}$ symmetry is restored during inflation and spontaneously broken at some epoch well after inflation. Then the network of cosmic strings and domain walls is formed, in which each string is attached by $4n$ domain walls. This network is stable if $V_\mathrm{QCD}$ were absent.
Due to $V_\mathrm{QCD}$, which works as a bias potential, the string-wall network finally collapses unless the two terms in $V_a$ have degenerate minima at $a \neq 0$.
In the following, we require that $4n$ and $k$ are coprime to each other so that the CP-conserving vacuum, $a = 0$, is finally realized in the whole universe.
Therefore, in this scenario, $V_{\rm QCD}$ takes the role of picking up a CP-conserving vacuum $a=0$ dynamically out of $4n$ vacua. We discuss the cosmological implications of such a scenario in the next section in detail.

On the other hand, if the $Z_{4n}$ symmetry is already broken during inflation, the field value of $\phi$ is effectively uniform at the end of inflation.
To solve the strong CP problem, $\phi$ must settle to $\theta_\phi = 0$, which is realized with a probability of $1/(4n)$ when we consider $V_\phi$.
If $V_\mathrm{QCD}$ is sufficiently larger than the tree level potential, $a$ finally settles to $a = 0$ and the strong CP problem is solved, although the Nelson-Barr mechanism is irrelevant.
In this case, the axion is trapped in a local minimum for a while and then starts to oscillate around the true vacuum as in the trapped misalignment mechanism~\cite{Higaki:2016yqk,Nakagawa:2020zjr,DiLuzio:2021gos,Jeong:2022kdr,Nakagawa:2022wwm,DiLuzio:2024fyt}.

\section{Cosmology of domain walls}
\label{sec: domain wall}

Here, we assume that $\phi$ acquires a nonzero VEV after inflation and discuss the evolution of the string-wall network focusing on its cosmological implications.%
\footnote{
    There are two types of domain walls: one is related to the breaking of $Z_4$ and the other $Z_{4n}$. The former is characterized by the VEV of $S$, and we assume that $S$ continues to have a large VEV as in the scenario of Ref.~\cite{Murai:2024alz}. Thus $S$ domain walls do not appear. We focus on the latter type of domain walls characterized by the VEV of $\phi$. 
}
In the following, we assume the radiation-dominated universe in all the relevant epoch.

In our scenario, after the formation of cosmic strings, domain walls are formed when the axion mass overcomes the Hubble friction.
Each cosmic string has $4n$ domain walls attached to it and forms a string-wall network.
The tensions of strings and walls are given as 
\begin{align}
    \mu
    \simeq
    \pi f_a^2
    \ , \quad 
    \sigma 
    \simeq 
    \frac{m_a f_a^2}{2 n^2}
    =
    \frac{4 \sqrt{\gamma_\phi} \epsilon^{2n+1}}{n} \Lambda^3
    \ ,
    \label{eq: tensions}
\end{align}
respectively.

In contrast to the case with a domain wall number of unity, the string-wall network is long-lived and eventually set in the so-called scaling regime.
In the scaling regime, there are $\mathcal{O}(1)$ strings and walls in each Hubble volume.
Thus, the energy densities of strings and walls evolve as
\begin{align}
    \rho_\mathrm{str} 
    \sim 
    \mu H^2 
    \ , \quad 
    \rho_\mathrm{wall} 
    \sim 
    \sigma H 
    \ ,
\end{align}
respectively.
Since $\rho_\mathrm{wall}$ decreases in time more slowly than $\rho_\mathrm{str}$, the energy density of the string-wall network is approximated by $\rho_\mathrm{wall}$.

As the universe cools down, the potential from the QCD effect grows and works as a potential bias on the domain walls.
The difference in the potential between two sides of a domain wall is evaluated as 
\begin{align}
    \Delta V_j
    &=
    |V_\mathrm{QCD}(a_{j})
    -
    V_\mathrm{QCD}(a_{j-1})|
    \nonumber \\
    &=
    \chi(T) \left|
        \cos\left( \frac{j k \pi}{2n} \right) 
        -
        \cos\left( \frac{(j-1) k \pi}{2n} \right) 
    \right|
    \ ,
\end{align}
where $a_j \equiv j \pi f_a/(2n)$ with $j = 0, \ldots, 4n$.
Note that $a_0$ and $a_{4n}$ represent the same vacuum.
Still there remains a degeneracy between $a_j$ and $a_{4n-j}$, which is ensured by the invariance under the CP transformation $\theta_\phi \to -\theta_\phi$. However, the lowest energy vacuum $a_0$ is unique.
In the following, we consider the case of $n \geq k$ and denote the typical bias by $\langle \Delta V \rangle \equiv k \chi(T)/n$, while the bias depends on $j$.%
\footnote{
    Precisely speaking, the potential minimum slightly deviates from $a_0=0$ as mentioned in Sec.~\ref{sec:modeldesc}, but the CP also transforms the phase of $S$ as $\theta_S \to -\theta_S$. Thus the degenerate potential minima are separated by the $S$ domain wall, which is irrelevant in our cosmological scenario.
}

The string-wall network collapses when this pressure overcomes the domain wall tension force, i.e., 
\begin{align}
    \rho_{\mathrm{wall}}(t_\mathrm{dec})
    =
    \frac{\mathcal{A} \sigma}{t_\mathrm{dec}}
    \simeq 
    \langle \Delta V \rangle
    \ ,
\end{align}
where $t_\mathrm{dec}$ is the cosmic time when the network collapses, and $\mathcal{A} = {\cal O}(1)$ is an area parameter representing the number of domain walls in each Hubble volume.
Although $\mathcal{A}$ evolves in time~\cite{Hiramatsu:2012sc,Hiramatsu:2013qaa} and depends on the initial conditions of $\phi$~\cite{Kitajima:2023kzu} in general, in the following, we set $\mathcal{A} = 1$ for simplicity.
Thus, we can evaluate $t_\mathrm{dec}$ as
\begin{align}
    t_\mathrm{dec}
    =
    \frac{1}{2H_\mathrm{dec}}
    \simeq 
    \frac{n \sigma}{k \chi(T_\mathrm{dec})}
    \ ,
\end{align}
where $T_\mathrm{dec}$ is the cosmic temperature at $t = t_\mathrm{dec}$.
From the relation during the radiation-dominated era, $H = 1/(2t)$, and the Friedmann equation, we obtain
\begin{align}
    T_\mathrm{dec}
    =
    \left\{
        \begin{array}{ll}
        \displaystyle
        \left(
            \frac{45 k^2}{2 \pi^2  n^2 g_*(T_\mathrm{dec})} 
            \frac{M_\mathrm{Pl}^2 \chi_0^2}{\sigma^2}
        \right)^{1/4}
        & 
        (T_\mathrm{dec} < T_\mathrm{QCD})
        \vspace{3mm}
        \\
        \displaystyle
        \left(
            \frac{45 k^2}{2 \pi^2  n^2 g_*(T_\mathrm{dec})} 
            \frac{M_\mathrm{Pl}^2 \chi_0^2 T_\mathrm{QCD}^{2p}}{\sigma^2}
        \right)^{1/(4+2p)}
        & 
        (T_\mathrm{dec} \geq T_\mathrm{QCD})
        \end{array}
    \right.
    \ .
\end{align}
For the domain wall network to collapse before the big bang nucleosynthesis, we require $T_\mathrm{dec} \gtrsim T_\mathrm{dec,min} = 10$\,MeV, which leads to the upper bound on $\sigma$ of
\begin{align}
    \sigma 
    \lesssim
    3.7 \times 10^{17}\,\mathrm{GeV}^3
    \times \frac{k}{n} 
    \left( \frac{g_{*,\mathrm{dec}}}{10.75} \right)^{-1/2}
    \left( \frac{T_\mathrm{dec,min}}{10\,\mathrm{MeV}} \right)^{-2}
\end{align}
We also require that the domain walls decay before they dominate the universe.
This condition leads to the upper bound of 
\begin{align}
    \sigma
    \lesssim 
    1.2 \times 10^{16}\,\mathrm{GeV}^3 \times \sqrt{k/n}
    \ .
    \label{eq: sigma upper bound from domination}
\end{align}
Note that the potential bias is at most $\sim \chi_0$, which is smaller than the total energy density at $T = T_\mathrm{QCD}$, and thus the domain walls collapse at $T_\mathrm{dec} < T_\mathrm{QCD}$ when $\sigma$ saturates this upper limit.

\subsection{Dark matter production}
\label{subsec: DM from DW}

First, we evaluate the dark matter abundance emitted from the domain wall collapse following Ref.~\cite{Kawasaki:2014sqa}.
Even in the scaling regime, the comoving number density of domain walls decreases emitting axion particles.
Since, in the scaling regime, the energy density of the domain walls decreases more slowly than non-relativistic matter, the axion abundance after the domain wall collapse is mainly contributed by particle production during the network collapse.
Here, we evaluate the axion abundance assuming that the axion particle is stable.

In the scaling regime during the radiation-dominated era, the domain walls emit axion particles at the rate of 
\begin{align}
    \left. 
        \frac{\mathrm{d}\rho_\mathrm{wall}}{\mathrm{d}t}
    \right|_\mathrm{emit}
    =
    -\frac{\sigma}{2 t^2}
    \ .
\end{align}
Note that this rate does not include the decrease in $\rho_\mathrm{wall}$ due to cosmic expansion, and we neglected the decrease due to gravitational wave emission, which is negligibly small.
We parameterize the average energy of emitted axions by
\begin{align}
    \bar{\omega}
    =
    \tilde{\epsilon} m_{a}
    \ ,
\end{align}
where we assume $\tilde{\epsilon}_a$ to be a constant.
Here, we also assume that the axion mass can be approximated by $m_a$ for any temperatures although $V_\mathrm{QCD}$ can substantially contribute to the axion mass in general as we will see later.
Then, the axion number density emitted from the domain walls is evaluated as
\begin{align}
    n_{\mathrm{dec}}(t)
    =
    \frac{\sigma}{\tilde{\epsilon} m_a t}
    \ ,
\end{align}
for $t < t_\mathrm{dec}$.
Thus, the energy density from the domain wall collapse is given by 
\begin{align}
    \rho_{\mathrm{dec}}(t)
    &\simeq 
    \frac{
    \sigma}{\tilde{\epsilon} t_\mathrm{dec}}
    \left( \frac{a(t_\mathrm{dec})}{a(t)} \right)^3
    \ ,
\end{align}
after the emitted axions become non-relativistic. 
Thus, we can evaluate the ratio of the energy density of $a$ to the entropy density $s$ as
\begin{align}
    \frac{\rho_{\mathrm{dec}}}{s}
    &\simeq 
    \frac{2\sigma H_\mathrm{dec}}{\tilde{\epsilon}}
    \frac{45}{4 \pi^2 g_{*s}(T_\mathrm{dec}) T_\mathrm{dec}^3}
    \nonumber \\
    &\simeq 
    \frac{45\sigma}{2 \pi^2 \tilde{\epsilon} g_{*s}(T_\mathrm{dec}) }
    \sqrt{\frac{\pi^2 g_*(T_\mathrm{dec})}{90}}
    \frac{1}{M_\mathrm{Pl} T_\mathrm{dec}}
    \nonumber \\
    &\simeq
    \frac{45\sigma}{2 \pi^2 \tilde{\epsilon} g_{*s}(T_\mathrm{dec}) M_\mathrm{Pl}}
    \sqrt{\frac{\pi^2 g_*(T_\mathrm{dec})}{90}}
    \left(
        \frac{2 \pi^2  n^2 g_*(T_\mathrm{dec})}{45 k^2} 
        \frac{\sigma^2}{M_\mathrm{Pl}^2 \chi_0^2 T_\mathrm{QCD}^{2p}}
    \right)^{1/(4+2p)}
    \nonumber \\
    &\simeq 
    0.41\,\mathrm{eV}
    \times \frac{1}{\tilde{\epsilon}}
    \left( \frac{n}{k} \right)^{\frac{1}{2+p}}
    \left( \frac{g_{*s,\mathrm{dec}}}{60} \right)^{-1}
    \left( \frac{g_{*,\mathrm{dec}}}{60} \right)^{\frac{3+p}{4+2p}}
    \left( \frac{\sigma}{5 \times 10^{9}\,\mathrm{GeV}^3} \right)^{\frac{3+p}{2+p}}
    \ .
    \label{eq: axion abundance}
\end{align}
for $T_\mathrm{dec} > T_\mathrm{QCD}$.
If $\rho_{\mathrm{dec}}/s = 0.44$\,eV (see, e.g., Ref.~\cite{Planck:2018vyg}), the axion emitted from the domain wall can explain all dark matter.

In addition to the contribution from the domain wall collapse, the axion is generated from the misalignment mechanism.
If we neglect $V_\mathrm{QCD}$, we can evaluate the axion abundance from the misalignment mechanism as~\cite{ParticleDataGroup:2024cfk}
\begin{align}
    \Omega_a^\mathrm{mis} h^2
    =
    0.12 \langle \theta_{\phi,\mathrm{i}}^2 \rangle
    \left( \frac{m_{a0}}{4.7 \times 10^{-19}\,\mathrm{eV}} \right)^{1/2}
    \left( \frac{f_{a}}{10^{16}\,\mathrm{GeV}} \right)^2
    \ ,
    \label{eq: misalignment}
\end{align}
where $h \equiv H_0/(100\,\mathrm{km/s/Mpc})$ is the reduced Hubble constant, and $\theta_{\phi,\mathrm{i}}$ is the initial value of $\theta_\phi$, and the angle bracket denotes the spatial average.
In our case, we obtain
\begin{align}
    \langle \theta_{\phi,\mathrm{i}}^2 \rangle
    =
    \frac{2n}{\pi} \int_{-\frac{\pi}{4n}}^{\frac{\pi}{4n}} \theta_{\phi,\mathrm{i}}^2 
    \mathrm{d}\theta_{\phi,\mathrm{i}}
    =
    \frac{\pi^2}{48n^2}
    \ .
\end{align}

\subsection{Gravitational wave production}
\label{subsec:GW from DW}

In the scaling regime, the string-wall network also emits gravitational waves.
Here, we evaluate the spectrum of the gravitational waves following Ref.~\cite{Hiramatsu:2013qaa}.
As in the case of the axion emission, the dominant contribution comes from just before {and during} the network collapse.
Since the frequency of the gravitational waves is determined by the Hubble scale at the emission, the peak frequency of the total gravitational spectrum is estimated by 
\begin{align}
    f_\mathrm{peak,dec}
    \sim
    H_\mathrm{dec}
\end{align}
at $t = t_\mathrm{dec}$.
The peak value of the density parameter of the gravitational waves, $\Omega_\mathrm{GW}$, is estimated as
\begin{align}
    \Omega_\mathrm{GW,dec}^\mathrm{peak}
    =
    \frac{\epsilon_\mathrm{GW} \sigma^2}{24\pi M_\mathrm{Pl}^4 H_\mathrm{dec}^2}
    \ ,
\end{align}
where $\epsilon_\mathrm{GW}$ is an efficiency parameter of the gravitational wave emission.

Taking into account the redshift after the emission, for $T_\mathrm{dec} < T_\mathrm{QCD}$, we obtain the peak frequency at the current time as
\begin{align}
    f_\mathrm{peak,0}
    &=
    \left( \frac{g_{*s,0}}{g_s(T_\mathrm{dec})} \right)^{1/3}
     \frac{T_0}{T_\mathrm{dec}}
    f_\mathrm{peak,dec}
    \nonumber \\
    &\sim
    14\,\mathrm{nHz} \times 
    \sqrt{\frac{k}{n}}
    \left( \frac{g_{*s,\mathrm{dec}}}{10.75} \right)^{-1/3}
    \left( \frac{g_{*,\mathrm{dec}}}{10.75} \right)^{1/4}
    \left( \frac{\sigma}{2.5 \times 10^{15}\,\mathrm{GeV}^3} \right)^{-1/2}
    \ ,
\end{align}
and the peak of the current gravitational wave spectrum as
\begin{align}
    \Omega_{\mathrm{GW},0}^\mathrm{peak} h^2
    &=
    \Omega_\mathrm{r,0} h^2 
    \frac{g_{*,\mathrm{dec}}}{g_{*,0}}
    \left( \frac{g_{*s,0}}{g_{*s,\mathrm{dec}}} \right)^{4/3}
    \Omega_\mathrm{GW,dec}^\mathrm{peak}
    \nonumber \\
    &\simeq
    1.9 \times 10^{-9} \times 
    \epsilon_\mathrm{GW} 
    \left( \frac{n}{k} \right)^2
    \left( \frac{g_{*s,\mathrm{dec}}}{10.75} \right)^{-4/3}
    \left( \frac{g_{*,\mathrm{dec}}}{10.75} \right)
    \left( \frac{\sigma}{2.5 \times 10^{15}\,\mathrm{GeV}^3} \right)^{4}
    \ ,
\end{align}
where $T_0 \simeq 2.725$\,K is the current CMB temperature~\cite{Fixsen:2009ug} and $\Omega_{\mathrm{r},0} = 4.15 \times 10^{-5} h^{-2}$ is the current density parameter of radiation~\cite{Planck:2018vyg}.
According to numerical simulations, the gravitational wave spectrum around the peak follows $\Omega_{\mathrm{GW},0} \propto f^3$ for $f < f_\mathrm{peak,0}$ and $\propto f^{-1}$ for $f > f_\mathrm{peak,0}$~\cite{Hiramatsu:2013qaa}.
Interestingly, if the domain walls collapse around the QCD phase transition epoch thanks to the QCD potential bias, the peak frequency is predicted to be nHZ range, which is the range of pulsar timing array experiments~\cite{Moroi:2011be,Ferreira:2022zzo,Kitajima:2023cek}.

\subsection{Model predictions}
\label{subsec: predictions}

\begin{figure}[htbp]
    \centering
    \includegraphics[width=.68\textwidth]{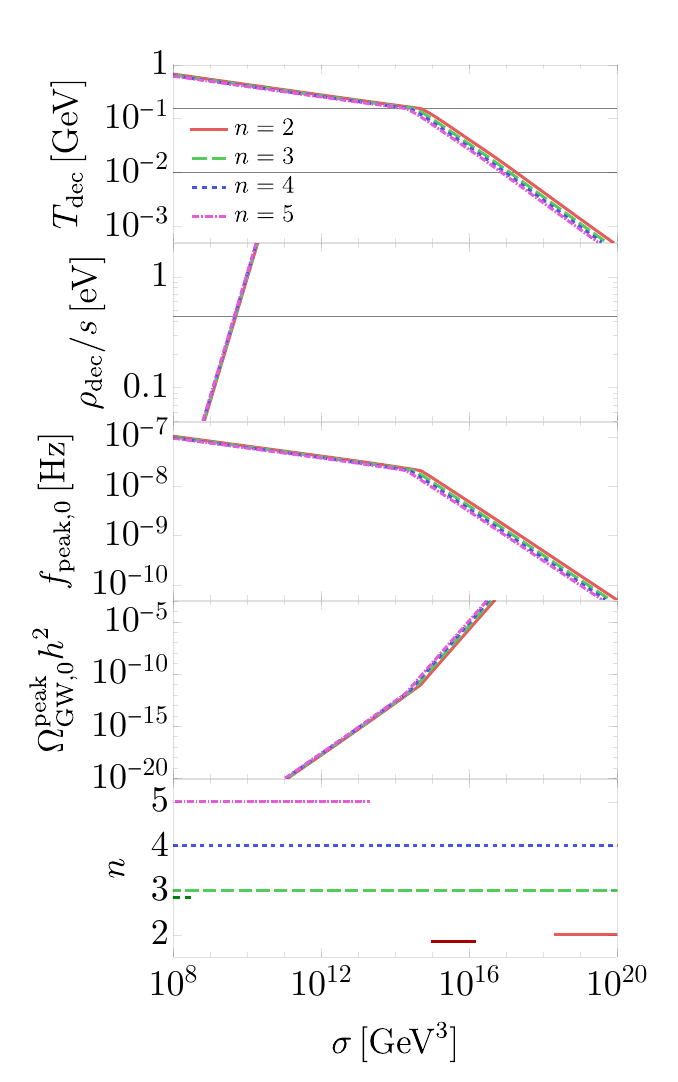}
    \caption{%
        Dependence of $T_\mathrm{dec}$, $\rho_\mathrm{dec}/s$, $f_\mathrm{peak,0}$, and $\Omega_\mathrm{GW,0}^\mathrm{peak}$ on $\sigma$.
        The red-solid, green-dashed, blue-dotted, and purple-dot-dashed lines correspond to $n = 2$, $3$, $4$, and $5$, respectively.
        The horizontal gray lines denote $T_\mathrm{dec} = T_\mathrm{QCD}$, $T_\mathrm{dec} = 10$\,MeV, and $\rho_\mathrm{dec}/s = 0.44$\,eV from top to bottom.
        The bottom plot shows the allowed region for $\sigma$ under the limit on $\epsilon$ in Eq.~\eqref{eq: epsilon range}.
        We use $\Lambda = M_\mathrm{Pl}$ and $\Lambda = 5 \times 10^{11}$\,GeV for the lighter and darker lines, respectively.
        We assume $k = 1$ and $\gamma_\phi = 0.1$ for all lines.
    }
    \label{fig: sigma dependnce}
\end{figure}
As seen in the previous section, the production of the axion and gravitational waves is mainly controlled by the domain wall tension, $\sigma$.
We show the dependence of $T_\mathrm{dec}$, $\rho_\mathrm{dec}/s$, $f_\mathrm{peak,0}$, and $\Omega_\mathrm{GW,0}^\mathrm{peak}$ on $\sigma$ in Fig.~\ref{fig: sigma dependnce}.
In the bottom plot, we also show the allowed region for $\sigma$ under the limit on $\epsilon$ in Eq.~\eqref{eq: epsilon range}.
Here, we adopted the fitting function of $g_*(T)$ and $g_{*s}(T)$ presented in Ref.~\cite{Saikawa:2018rcs}.
While all dark matter can be produced for $\sigma \simeq 5 \times 10^9\,\mathrm{GeV}^3$, the abundance of the gravitational waves is much smaller than the observable range for such $\sigma$.

Once we fix $n$ and $\Lambda$, we can relate $\sigma$ to $\epsilon$ via Eq.~\eqref{eq: tensions}, leading to $m_a$ and $f_a$.
Then, we can estimate the mass and couplings of the axion.
The axion mass is determined by the two contributions in $V_a$ (see Eq.~\eqref{eq: axion potential}).
For $T<T_\mathrm{QCD}$, the axion mass squared is given by 
\begin{align}
    m_{a0}^2
    \equiv
    \frac{k^2 \chi_0}{f_a^2} 
    + m_a^2
    \ .
\end{align}
Regarding the couplings of the axion, we expect that the axion is coupled to the standard model particles as the QCD axion.
In particular, we assume the axion-photon coupling of 
\begin{align}
    \mathcal{L}
    \supset 
    - \frac{g_{a\gamma\gamma}}{4} a F_{\mu \nu} \tilde{F}^{\mu \nu}
    \ ,
\end{align}
with 
\begin{align}
    g_{a\gamma\gamma}
    \sim
    \frac{k^2 \alpha}{2\pi f_a}
    \ .
\end{align}
Here, $F_{\mu \nu}$ and $\tilde{F}^{\mu\nu}$ are the field strength of the photon and its dual, and $\alpha \simeq 1/137$ is the fine-structure constant.

\begin{figure}[htbp]
    \centering
    \includegraphics[width=.8\textwidth]{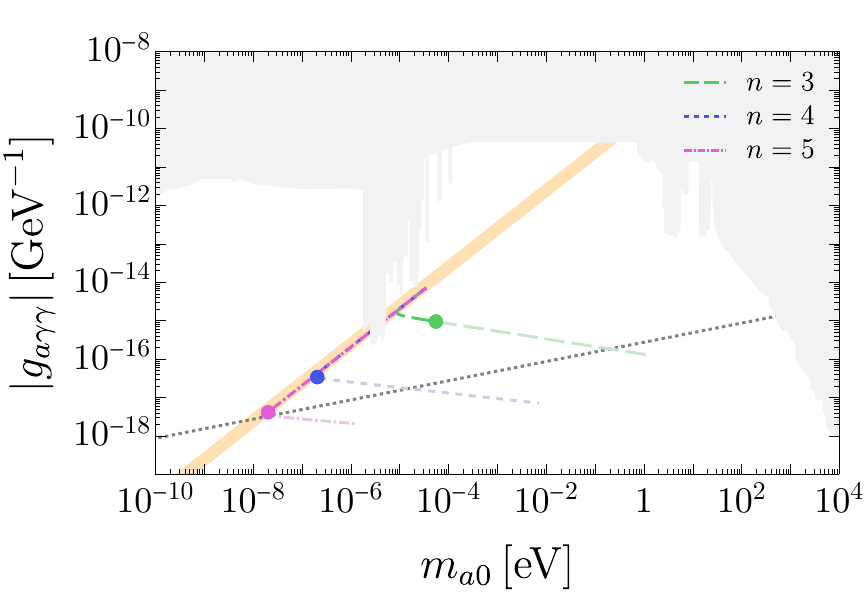}
    \caption{%
    Mass and photon coupling of the axion for $\Lambda = M_\mathrm{Pl}$.
    The colored lines correspond to different values of $n$ as in Fig.~\ref{fig: sigma dependnce}.
    The colored circles correspond to the parameter predicting all dark matter, $\rho_\mathrm{dec}/s = 0.44$\,eV, from the domain wall collapse.
    On the heavier side of the circles, shown in lighter colors, such contribution is overabundant.
    The orange band is the theoretical prediction for the QCD axion models.
    The gray region denotes the current constraints on the axion.
    We adopted the experimental constraints from Ref.~\cite{AxionLimits}.
    On the gray-dotted line, the axions produced from the misalignment mechanism~\eqref{eq: misalignment} match the observed abundance of dark matter.
    Here, we used $k = 1$, $\tilde{\epsilon} = 1$, and $\gamma_\phi = 0.1$.
    }
    \label{fig: ma vs g}
\end{figure}

First, we fix $\Lambda = M_\mathrm{Pl}$ and discuss the production of axion dark matter.
We show the prediction for $(m_{a0},g_{a\gamma\gamma})$ in this case in Fig.~\ref{fig: ma vs g}.
The colored lines denote the prediction of this scenario.
Due to the contribution of $V_\phi$, the axion can be heavier than the QCD band.
On the other hand, when $m_a$ is smaller than the contribution from $V_\mathrm{QCD}$, the prediction is aligned with the QCD band.
At the colored circles, the axions emitted from the domain wall collapse can account for all dark matter.
For these parameters, the contribution of $V_\mathrm{QCD}$ to the axion mass is not dominant, and thus the temperature dependence of the axion mass does not affect the estimate of the axion abundance significantly.
Note that on the heavier side of the circles, shown in lighter colors, the produced axions exceed the dark matter abundance, and thus it should decay via some processes, although our model does not include such decay channels as it is.
The gray-dotted line represents the parameter with which the contribution from the misalignment mechanism~\eqref{eq: misalignment} matches the observed abundance of dark matter.
While the contributions from the domain wall collapse and misalignment mechanism can comparably contribute to all dark matter for $n =5$, the former contribution will be dominant for $n = 3$ and $4$.%
\footnote{
    Here we have implicitly assumed a cosmological scenario with domain wall formation. If the symmetry is already broken during inflation, there are no domain walls at all. In such a case, the misalignment mechanism is the dominant dark matter production mechanism.
}

\begin{figure}[!t]
    \begin{center}  
        \includegraphics[width=0.8\textwidth]{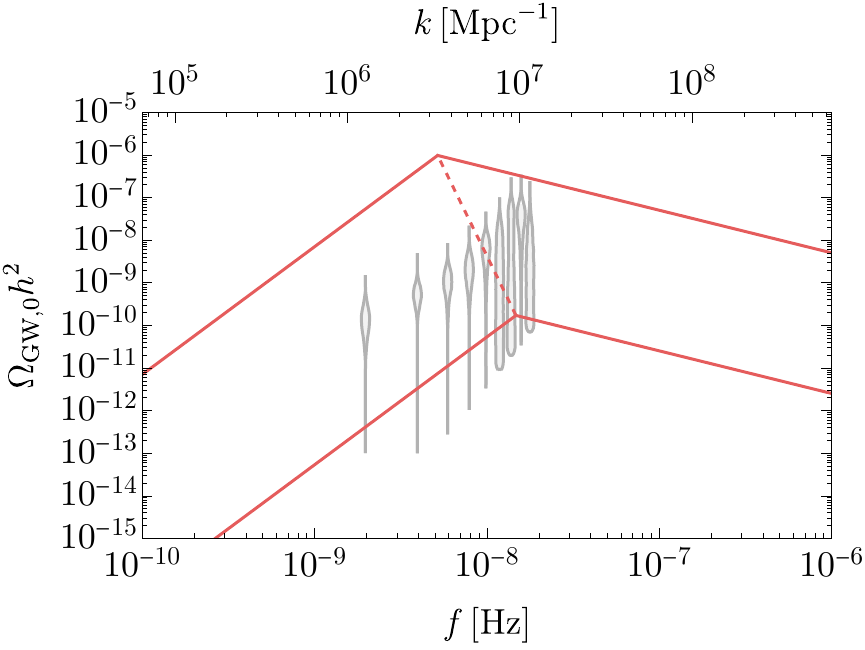}
        \end{center}
    \caption{%
        Current power spectrum of the gravitational waves from the domain wall collapse.
        The red-solid lines denote the maximum and minimum power spectrum of the gravitational waves.
        The red-dashed line denotes the contour of the peak of the spectrum.
        The gray violins are from the NANOGrav 15\,yr result~\cite{NANOGrav:2023gor}.
        Here, we use $\Lambda = 5 \times 10^{11}$\,GeV, $n = 2$, $k = 1$, $\epsilon_\mathrm{GW} = 1$, and $\gamma_\phi = 0.1$.
    }
    \label{fig: GW} 
\end{figure}

\begin{figure}[!t]
    \begin{center}  
        \includegraphics[width=0.8\textwidth]{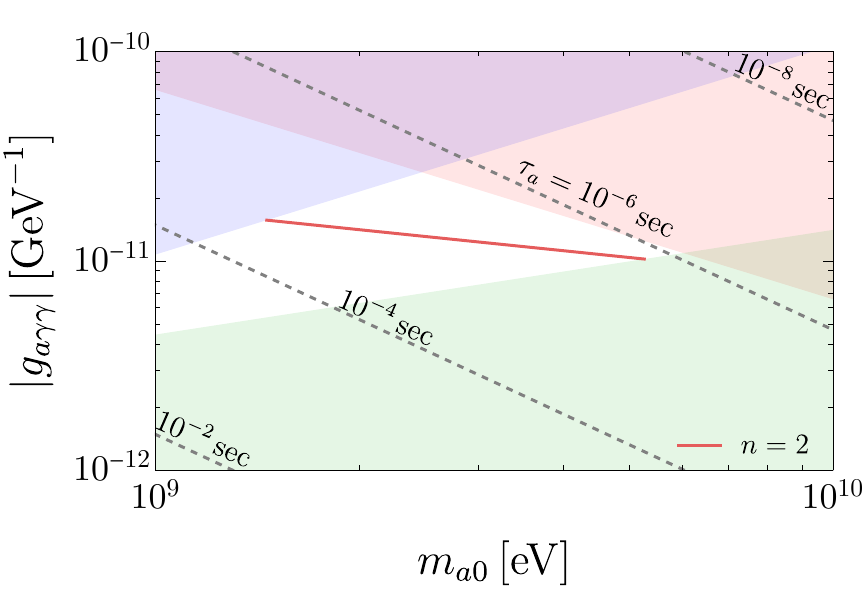}
        \end{center}
    \caption{%
        The red line represents the mass and photon coupling of the axion for $\Lambda = 5 \times 10^{11}$\,GeV and $n = 2$.
        The gray-dashed lines represent the contours of the axion lifetime, $\tau_a$.
        The shaded regions represent the limited parameters when we vary $\Lambda$ and $f_a$.
        The red and blue regions correspond to the limits on $\epsilon$ shown in Fig.~\ref{fig: epsilon range}.
        In the green region, the domain walls dominate the universe before the decay.
        Here, we used $k = 1$ and $\gamma_\phi = 0.1$.
    }
    \label{fig: ma vs g heavy} 
\end{figure}

Next, we consider $\Lambda = 5 \times 10^{11}$\,GeV and discuss the generation of gravitational waves.
We show the current power spectrum of the gravitational waves from the domain wall collapse in Fig.~\ref{fig: GW}.
Here, we assume $n = 2$, $k = 1$, $\epsilon_\mathrm{GW} = 1$, and $\gamma_\phi = 0.1$.
The red lines denote the maximum and minimum power spectrum of the gravitational waves coming from the bound on $\epsilon$~\eqref{eq: epsilon range} and the requirement for the domain wall collapse before domination~\eqref{eq: sigma upper bound from domination}.
Within this range, the gravitational waves from the domain wall collapse can explain the signal reported by the NANOGrav 15\,yr result~\cite{NANOGrav:2023gor}.%
\footnote{
    The explanation of the NANOGrav result by the domain wall collapse due to the QCD potential is also numerically studied in Ref.~\cite{Kitajima:2023cek}.
}
We also show the corresponding range of the mass and photon coupling in Fig.~\ref{fig: ma vs g heavy}.
The gray-dashed lines represent the contours of the axion lifetime, $\tau_a$.
The decay rate of the heavy axion with $m_{a0} > 1$\,GeV is estimated by
\begin{align}
    \Gamma_a
    =
    \tau_a^{-1}
    \simeq 
    \frac{k^2 \alpha_s^2}{32\pi^3} \frac{m_{a0}^3}{f_a^2}
    \ ,
\end{align}
where $\alpha_s$ is the strong coupling constant.
Here, we adopt $\alpha_s = 0.2$ as a typical value for the scale of $1\,\text{--}\,10$\,GeV~\cite{dEnterria:2022hzv}.
The axion lifetime for the parameters of interest is shorter than $10^{-4}$\,sec, which corresponds to the Hubble time at the domain wall collapse, $T = \mathcal{O}(0.1)$\,GeV (see Fig.~\ref{fig: sigma dependnce}).
Thus, the axions will immediately decay after emitted from the domain wall collapse.
In the colored regions, $\bar{\theta}_s$ is too large (red), the CKM matrix cannot be reproduced (blue), and the domain walls dominate the universe before the decay (green).
This shows that the parameter region for reproducing the NANOGrav result may be verified by the future measurement of the strong CP angle or the non-unitarity of the CKM matrix (or the collider production of heavy quarks).

\section{Conclusions and discussion}
\label{sec: conclusion}

The BBP model~\cite{Bento:1991ez} is a simple model that solves the strong CP problem through the Nelson-Barr mechanism, although there are several quality problems in the original setup.
A modified BBP model proposed in Ref.~\cite{Murai:2024alz} introduced an approximate $Z_{4n}$ symmetry to suppress several operators that cause quality problems. There a small explicit symmetry breaking parameter $\epsilon$ was introduced by hand. 
In this paper, we have considered a concrete realization of such a setup by introducing a complex scalar $\phi$ and assuming a spontaneous breaking of the $Z_{4n}$ symmetry due to the VEV of $\phi$, rather than the explicit breaking.
The small symmetry breaking parameter is replaced by the ratio of the VEV of $\phi$ and cutoff scale $\Lambda$.

We have studied the cosmology and phenomenology of such a model. 
First, the phase of $\phi$, $\theta_\phi$, is relatively light due to the $Z_{4n}$ symmetry, and it can play the role of dark matter depending on the parameter choice. 
Interestingly, the phase of $\phi$ also obtains a mass from the QCD instanton effect in the same way as the QCD axion since $\phi$ necessarily couples to heavy quarks.
In our model, the strong CP problem is solved even if the QCD instanton effect is much smaller than the tree-level potential.
Still, the phase of $\phi$ shares several properties similar to the QCD axion, and hence we call it the Nelson-Barr axion.
Compared with the conventional QCD axion, the Nelson-Barr axion can take different parameter ranges in terms of the mass and coupling strength (see Figs.~\ref{fig: ma vs g} and \ref{fig: ma vs g heavy}). 

Second, domain walls lead to several important consequences.
In the present model, domain walls associated with the spontaneous breakdown of the $Z_{4n}$ symmetry are not stable due to the potential bias induced by the QCD instanton effect~\cite{Preskill:1991kd}.
Thus domain walls eventually collapse and, as a result, the whole universe dynamically ends up with the CP-conserving vacuum $\theta_\phi=0$. 
The phenomenological implications of domain wall collapse depend on the cutoff scale $\Lambda$.
If $\Lambda$ is huge (say, close to the Planck scale), it can produce the correct amount of Nelson-Barr axion dark matter.
On the other hand, if $\Lambda$ is as small as $10^{12}\,$GeV, it can explain the recent pulsar timing array data for the stochastic gravitational wave background. 
Note that, in the most previous models of collapsing domain walls with the QCD potential bias, the strong CP problem is a serious issue since the Peccei-Quinn mechanism does not work, and some non-trivial model extensions are required~\cite{Preskill:1991kd,Dine:1993yw,Riva:2010jm,Moroi:2011be,Hamaguchi:2011nm,Higaki:2016yqk,Higaki:2016jjh,Chigusa:2018hhl,Chiang:2020aui,Ferreira:2022zzo,Kitajima:2023cek,Bai:2023cqj,Blasi:2023sej}.
We emphasize that our model provides a natural realization of such a scenario {\it without} the strong CP problem; even if the QCD-induced potential for the Nelson-Barr axion is subdominant, the potential minimum is a CP-conserving point.
Interestingly, the parameter region for reproducing the NANOGrav result is close to the current experimental bound from the measurement of the strong CP angle and the unitarity of the CKM matrix, as shown in Fig.~\ref{fig: ma vs g heavy}. In other words, future improvements of these observables may verify or falsify this scenario.

\section*{Acknowledgment}

This work was supported by World Premier International Research Center Initiative (WPI), MEXT, Japan.
This work was also supported by JSPS KAKENHI (Grant Numbers 24K07010 [KN], 23KJ0088 [KM], and 24K17039 [KM]).

\bibliographystyle{utphys}
\bibliography{ref}

\end{document}